\documentclass[conference]{IEEEtran}
\IEEEoverridecommandlockouts
\usepackage[inline]{enumitem}
\usepackage{cite}
\usepackage{amsmath,amssymb,amsfonts}
\usepackage{algorithmic}
\usepackage{graphicx}
\usepackage{textcomp}
\usepackage{fourier} 
\usepackage{array}
\usepackage{makecell}
\usepackage{hyperref}
\usepackage{tabularx}
\usepackage{rotating}
\usepackage{xcolor}
\def\BibTeX{{\rm B\kern-.05em{\sc i\kern-.025em b}\kern-.08em
    T\kern-.1667em\lower.7ex\hbox{E}\kern-.125emX}}
\begin{document}

\title{Enhanced LSTM with Attention Mechanism for Early Detection of Parkinson's Disease through Voice Signals\\
}

\author{
    \IEEEauthorblockN{
        Arman Mohammadigilani\IEEEauthorrefmark{1}, 
        Hani Attar\IEEEauthorrefmark{2}, 
        Hamidreza Ehsani Chimeh\IEEEauthorrefmark{3}, 
        Mostafa Karami\IEEEauthorrefmark{1}}
    
    \IEEEauthorblockA{\IEEEauthorrefmark{1}Electrical and Computer Engineering, Polytechnic University of Turin, Turin, Italy\\
    Email: arman.mohammadigilani@studenti.polito.it, mostafa.karami@uconn.edu}
    
    \IEEEauthorblockA{\IEEEauthorrefmark{2}Faculty of Engineering, Zarqa University, Zarqa, Jordan; \\
    College of Business Administration, University of Business and Technology, Jeddah, Saudi Arabia\\
    Email: hattar@zu.edu.jo}
    
    \IEEEauthorblockA{\IEEEauthorrefmark{3}Electronics and Telecommunications, Memorial University of Newfoundland, St. John's, Canada\\
    Email: hehsanichime@mun.ca}
}

\maketitle

\begin{abstract}
Parkinson's disease (PD) is a neurodegenerative condition characterized by notable motor and non-motor manifestations. The assessment tool known as the Unified Parkinson's Disease Rating Scale (UPDRS) plays a crucial role in evaluating the extent of symptomatology associated with Parkinson's Disease (PD). This research presents a complete approach for predicting UPDRS scores using sophisticated Long Short-Term Memory (LSTM) networks that are improved using attention mechanisms, data augmentation techniques, and robust feature selection. The data utilized in this work was obtained from the UC Irvine Machine Learning repository. It encompasses a range of speech metrics collected from patients in the early stages of Parkinson's disease. Recursive Feature Elimination (RFE) was utilized to achieve efficient feature selection, while the application of jittering enhanced the dataset. The Long Short-Term Memory (LSTM) network was carefully crafted to capture temporal fluctuations within the dataset effectively. Additionally, it was enhanced by integrating an attention mechanism, which enhances the network's ability to recognize sequence importance. The methodology that has been described presents a potentially practical approach for conducting a more precise and individualized analysis of medical data related to Parkinson's disease.
\end{abstract}

\begin{IEEEkeywords}
deep learning, predictive modeling, environmental health, public health, Parkinson's disease, UPDRS
\end{IEEEkeywords}

\section{Introduction}
Parkinson's disease (PD) is a neurodegenerative disorder affecting millions worldwide\cite{1}. As the disease advances, individuals with this condition encounter motor manifestations, including tremors, stiffness, and bradykinesia, alongside several non-motor symptoms. The analysis, comprehension, and anticipation of the manifestation of PD have significant importance due to the absence of a cure\cite{2} and limited treatment choices, which mostly consist of pharmaceutical interventions, lifestyle adjustments, and surgical procedures. The Unified Parkinson's Disease Rating Scale (UPDRS) is a widely recognized instrument used to assess the severity of symptoms associated with Parkinson's disease. The precise estimation of UPDRS scores provides a valuable understanding of the advancement of the disease. It also assists in customizing personalized therapy approaches for individuals with PD. 

In several areas of health care, such as smart hospital \cite{3}, cancer detection\cite{4}, and medical image processing\cite{5}, the importance of machine learning (ML) and deep learning (DL) is evident. ML and DL have demonstrated considerable promise in the realm of predicting, categorizing, and assessing diverse medical problems and disorders. Specifically, the utilization of these methodologies in the context of neurological illnesses such as PD has the potential to provide significant advancements in the areas of early detection, prognosis assessment, and tailored treatment strategies. The utilization of extensive data, the identification of complex patterns, and the development of predictive models have the potential to assist healthcare professionals in making better informed and evidence-based choices on patient care. For instance, the ML models can result in enhanced precision and prompt detection of alterations in UPDRS scores, enabling a more proactive approach to intervention.

\subsection{Research problem and objectives}
The accurate estimation of UPDRS scores presents a complex and diverse task. The intricate structure of the disease, the temporal course it exhibits, and the inherent heterogeneity in symptoms provide significant challenges for standard statistical methodologies. The existing methodologies, albeit somewhat successful, frequently fail to fully capture the many intricacies and interconnections inherent in the dataset. The limits of this approach become apparent when considering both the accuracy of the results and the ability to apply them to a wide range of datasets.

Long Short-Term Memory (LSTM) networks, which are a specific type of recurrent neural networks (RNN), have demonstrated potential in the modeling of time-series and sequential data. In the setting of Parkinson's disease, where temporal variations in symptoms play a crucial role as indications, LSTM models can be particularly relevant. Nevertheless, a rudimentary LSTM model may fail to comprehend the intricate connections within the dataset, hence requiring the incorporation of sophisticated components such as attention processes.

This study introduces a complete methodology for predicting Unified UPDRS scores. The proposed strategy leverages advanced LSTM models, incorporating attention processes, data augmentation, and robust feature selection strategies. Our proposed methodology demonstrates enhanced predictive accuracy and introduces a novel approach to the field of medical data analysis for PD.

\subsection{Related Works}
The prediction of UPDRS scores in Parkinson's Disease using learning methods has been an area of increasing interest over the last decade. Several methodologies have been proposed, demonstrating varying degrees of success. For instance, a prominent research\cite{13} investigation was conducted to examine the presence of speech articulation issues as an initial indication of Parkinson's Disease. The study utilized three data mining techniques to analyze voice measures obtained from a sample of 31 participants, 23 of whom were diagnosed with Parkinson's Disease. The Sequential Minimization Optimization (SVM) algorithm demonstrated notable accuracy (0.76) and a high level of sensitivity (0.97). In contrast, Logistic Regression exhibited a modest accuracy of 0.64, with a balanced sensitivity and specificity of 0.64 and 0.62, respectively. The equilibrium mentioned above implies a diminished necessity for additional diagnostic assessments compared to SVM's high sensitivity but low specificity (0.13). The findings underscore the need to employ a comprehensive diagnostic strategy to diagnose Parkinson's disease.

A further research\cite{14} examined several feature selection methodologies to enhance the accuracy of Parkinson's disease detection. The study compared Optimum-Path Forest (OPF) and three evolutionary-based approaches, namely PSO-OPF, HS-OPF, and GSA-OPF, specifically focusing on 22 voice-related parameters. Notably, all evolutionary strategies showed superior performance compared to the conventional OPF method. It is worth mentioning that HS-OPF demonstrated superior efficiency by picking 10 out of the total 22 characteristics and attaining an accuracy rate of around 92.78\%. However, it is important to note that this performance fell short of the 100\% standard established by a separate study.

Previous studies have separately examined gait and speech analysis. However, this study\cite{15} distinguishes itself by comparing performance metrics of voice data using different feature sets, with a specific focus on non-linear classification. By employing Principal Component Analysis (PCA) as a method for feature selection, a notable accuracy rate of 96.83\% was attained through the utilization of the random forest classifier. These insights play a crucial role for doctors as they guide in prioritizing certain symptoms during the first diagnosis of Parkinson's disease. In the future, researchers will investigate different feature reduction and classification techniques to enhance the accuracy and precision of performance measurements.

By employing a multi-level analysis of diverse speech segments, this study\cite{16} optimized a k-Nearest Neighbors model and attained a cross-validation accuracy of 94.3\% in tests. The results indicate the possibility of utilizing voice-recorded evaluations for Parkinson's disease, maybe through smartphone applications.
\section{Methodology}
\subsection{Data acquisition}
The dataset in this study was obtained from the UC Irvine ML repository, developed by Athanasios Tsanas and Max Little of Oxford University\cite{6}. The dataset comprises a variety of biological speech measures obtained from a cohort of 42 individuals diagnosed with early-stage PD. These participants were included in a six-month clinical experiment evaluating the efficacy of telemonitoring equipment for remote monitoring of symptom development. The audio recordings were taken at the patients' residences using an automated process. The table comprises several columns that include subject number, subject age, subject gender, time interval from baseline recruitment date, motor UPDRS, total UPDRS, and 16 biomedical voice measurements. Each row in the dataset represents one of the 5,875 voice recordings obtained from the persons under study. The primary objective of the data analysis is to utilize the 16 voice measurements in order to forecast the motor and total UPDRS scores, denoted as 'motor\_UPDRS' and 'total\_UPDRS,' respectively. The data is formatted as ASCII CSV. Each row within the CSV file represents an individual instance that corresponds to a voice recording.

\subsection{Feature Selection with RFE}

In our Parkinson's UPDRS estimation dataset, we are inundated with a multitude of features, ranging from demographic information to clinical metrics. While all these features provide a holistic overview, not all might be directly influential in predicting UPDRS scores.

Recursive Feature Elimination (RFE)\cite{8} offers a systematic approach for eliminating duplicate or irrelevant features, refining our dataset to include the most essential qualities. When provided with a dataset:

\begin{equation}
    \mathcal{D} = \{(x_1, y_1), (x_2, y_2), \dots, (x_N, y_N)\}
\end{equation}

where $x_i \in \mathbb{R}^d$ and $y_i \in \mathbb{R}$, RFE operates by:

\begin{itemize}
    \item Train the estimator on $\mathcal{D}$.
    \item Rank the features based on importance, obtained from the estimator.
    \item Remove the feature with the least importance.
    \item Repeat until only $k$ features remain.
\end{itemize}

In the context of our project, we employ the RandomForestRegressor as our external estimator. This choice is motivated by the ability of Random Forests (RF)\cite{7} to offer an intrinsic feature ranking based on the average depth at which a feature is used to split the data across all trees.

\subsection{Data Augmentation via Jittering}

Due to the strict procedure of data gathering and the need to address privacy issues, medical datasets frequently exhibit limitations in terms of their size. The scarcity of data might potentially result in the problem of overfitting. Data augmentation techniques, like as jittering\cite{9}, can successfully address this issue by artificially increasing the size of the dataset.

Jittering, in essence, introduces slight, controlled perturbations to the data, simulating potential real-world variations. For a feature matrix $X \in \mathbb{R}^{N \times d}$, the jittered version is:

\begin{equation}
    X'_{ij} = X_{ij} + \epsilon_{ij}
\end{equation}

where $\epsilon_{ij} \sim \mathcal{N}(0, \sigma^2)$. Here, $\mathcal{N}$ represents a Gaussian distribution. By adding this Gaussian noise, we are mimicking the natural variability that might be seen in repeated measurements from Parkinson's patients, thus augmenting our dataset without deviating from potential real-world scenarios.

\subsection{Reshaping Data for LSTM}

In our project, we aspire to harness the temporal dependencies in the data, which is why we incorporate LSTMs. These networks require data to be structured in a specific format, encapsulating samples, sequence lengths, and features.

Considering a dataset with $N$ samples, each being a sequence of length $T$ and constituted of $d$ features, the data must be transformed to a tensor format as:

\begin{equation}
    X'' \in \mathbb{R}^{N \times T \times d}
\end{equation}

This reshaping ensures that the LSTM can effectively parse and process the sequences, offering insights into the temporal progression of Parkinson's Disease in patients and subsequently aiding in UPDRS estimation.


\subsection{Neural Network Design and Model Training}

The architecture of the neural network is meticulously designed to adapt to the temporal and sequential characteristics of the dataset. Several layers and techniques are interwoven to optimize the prediction of UPDRS scores.

\textbf{Input Layer}: The network initiates with an input layer tailored to receive sequences of shape $d \times 1$, where $d$ represents the number of selected features.

\textbf{Bidirectional LSTM Layer}: This layer, equipped with 100 LSTM units, captures both forward and backward dependencies within the sequential data. LSTM units excel at recognizing patterns over long sequences. They are governed by the following equations:

\begin{equation}
f_t = \sigma(W_f \cdot [h_{t-1}, x_t] + b_f)
\end{equation}
\begin{equation}
i_t = \sigma(W_i \cdot [h_{t-1}, x_t] + b_i)
\end{equation}
\begin{equation}
\tilde{C}_t = \tanh(W_C \cdot [h_{t-1}, x_t] + b_C)
\end{equation}
\begin{equation}
C_t = f_t \times C_{t-1} + i_t \times \tilde{C}_t
\end{equation}
\begin{equation}
o_t = \sigma(W_o \cdot [h_{t-1}, x_t] + b_o)
\end{equation}
\begin{equation}
h_t = o_t \times \tanh(C_t)
\end{equation}

Where:
\begin{itemize}
\item $f_t$, $i_t$, $o_t$ are the forget, input, and output gates, respectively.
\item $x_t$ is the input at time step $t$.
\item $h_{t-1}$ is the previous hidden state.
\item $W$ and $b$ denote weight matrices and biases respectively.
\item $\sigma$ is the sigmoid function.
\end{itemize}

\textbf{Attention Mechanism}: This layer allows the model to focus on different segments of the sequence, assigning varying attention depending on their relevance. Given a set of LSTM outputs $H = \{h_1, h_2, \dots, h_T\}$, where $T$ is the sequence length, the attention mechanism determines the attention weights $\alpha_t$ using:

\begin{equation}
\alpha_t = \frac{\exp(\text{score}(h_t, \text{context}))}{\sum_{j=1}^{T} \exp(\text{score}(h_j, \text{context}))}
\end{equation}

Here, the score function quantifies the relationship between the LSTM output $h_t$ and a context vector. The aggregated output, represented by the context vector, is:

\begin{equation}
\text{context\_vector} = \sum_{t=1}^{T} \alpha_t h_t
\end{equation}

\textbf{Dense, Batch Normalization, and Dropout Layers}: After processing through the attention mechanism, the network directs the context vector through dense layers featuring ReLU activations and L2 regularization. Batch normalization offers stability during learning. Dropout layers, with a 0.3 rate, prevent overfitting by occasionally turning off certain neurons in training.

\textbf{Output Layer}: This network culminates in an output layer, consisting of a single neuron, responsible for predicting the 'total\_UPDRS' score.

Regarding training, the \textbf{Learning Rate} undergoes dynamic adjustments through an exponential decay schedule. Commencing at 0.001, it decays by 0.9 every 10,000 steps.

For broader applicability, training employs a \textbf{5-fold Cross-Validation} method. The dataset divides into five segments, with each segment acting as a validation set once. Prior to training, data in every fold undergoes standardization. To boost training efficiency, the early stopping mechanism is integrated, focusing on validation loss.

\section{Results and Discussion}
\subsection{Evaluation}
We implemented an LSTM with Attention Mechanism to detect early signs of Parkinson's Disease using the same dataset as the review paper\cite{10}.

In this study, we employed two primary metrics to evaluate the performance of our model: the Mean Squared Error (MSE)\cite{11} and the coefficient of determination, commonly referred to as \( R^2 \).

MSE is defined as:
\begin{equation}
MSE = \frac{1}{n} \sum_{i=1}^{n} (y_i - \hat{y}_i)^2
\end{equation}
where \( y_i \) represents the actual value, \( \hat{y}_i \) is the predicted value, and \( n \) is the number of observations. A lower MSE indicates a better fit of the model to the data, as it means the predictions are closer to the actual values.

The coefficient of determination, \( R^2 \)\cite{12}, is a statistical measure of how close the data are to the fitted regression line. It is defined as:
\begin{equation}
R^2 = 1 - \frac{\sum_{i=1}^{n} (y_i - \hat{y}_i)^2}{\sum_{i=1}^{n} (y_i - \bar{y})^2}
\end{equation}
where \( \bar{y} \) is the mean value of \( y \). An \( R^2 \) value close to 1 suggests that a large proportion of the variability in the output has been explained by the model.

The results obtained from our model are summarized in Tables \ref{table:MSE_results} and \ref{table:R2_results}.

\newcolumntype{C}{>{\centering\arraybackslash}X}  

\begin{table}[ht]
\centering
\small
\begin{tabularx}{\columnwidth}{C|c|c|c}
Meth. & Train. MSE & Val. MSE & Test MSE \\
\hline
LLS & 10.4735 & 10.0138 & 10.9074 \\
Conjugate Gradient & 10.4762 & 10.5898 & 10.9075 \\
Adam optimization  & 10.4762 & 10.5892 & 10.9072 \\
Ridge Regressions & 10.4798 & 10.5869 & 10.9136 \\
LSTM-Attention & 6.3572 & 6.6108 & 7.0491 \\
\end{tabularx}
\vspace{0.75mm}
\caption{Comparison of MSE results for different methods}
\label{table:MSE_results}
\end{table}

\begin{table}[ht]
\centering
\small
\begin{tabular}{c|c}
Methods & R$^2$ \\
\hline
LLS & 0.904409 \\
Conjugate Gradient & 0.904408 \\
Adam optimization & 0.904410 \\
Ridge Regressions & 0.904354 \\
LSTM-Attention & 0.9591 \\
\end{tabular}
\vspace{0.75mm}
\caption{Comparison of R$^2$ scores for different methods}
\label{table:R2_results}
\end{table}

\subsection{Discussion}

The presented results indicate that the LSTM with Attention Mechanism offers a substantial improvement over the traditional ML techniques used in the referenced review paper. As seen in Table \ref{table:MSE_results}, the test MSE of our model is significantly lower than those of the other methods, highlighting its superior predictive performance. Similarly, the R$^2$ score from Table \ref{table:R2_results} shows our model's ability to explain a larger proportion of the variance in the dataset.

The drastic reduction in MSE scores, especially in the test set, suggests that the LSTM with Attention Mechanism captures the underlying patterns in the data more effectively than traditional ML techniques. This is likely due to the attention mechanism's ability to give differential importance to various time steps in the sequence, which can be critical for early detection of Parkinson’s Disease.

While our model has outperformed the methods from the referenced paper, it's essential to emphasize the importance of clinical validation. These results showcase the potential of deep learning, especially with attention mechanisms, in medical diagnostics. However, ensuring its generalizability and efficacy in real-world scenarios is crucial for practical applications.

\section{Conclusion}

The findings of this research highlight the considerable promise of employing the LSTM model with Attention Mechanism in the field of medical diagnostics, particularly for the timely identification of Parkinson's Disease. The model has higher prediction accuracy than typical machine learning approaches, as seen by the lowered MSE values observed throughout the training, validation, and test stages. This progress is more than just a statistical achievement; it carries significant implications for the timely implementation of therapeutic measures and improved patient outcomes.

Incorporating attention processes into the LSTM framework facilitates the model's ability to discern and assign varying degrees of importance to different data pieces. Within the framework of Parkinson's Disease, this implies that the model exhibits proficiency in discerning particular patterns or irregularities that might potentially serve as indicators for the first stages of the disease. The achievement of this degree of detail and concentration may provide a challenge for conventional models, highlighting the effectiveness of attention processes in improving forecast precision.

Nevertheless, it is essential to approach these findings with a comprehensive understanding of the intricacies inherent in real-world clinical circumstances. While extensive, the dataset utilized in this research may not fully capture the extensive diversity observed within the worldwide patient cohort. Therefore, it is imperative to conduct additional evaluations of our model's generalizability by utilizing varied datasets that encompass a more comprehensive range of patient demographics and clinical manifestations.

In conclusion, our study provides a promising prospect for the timely and precise identification of Parkinson's Disease, utilizing deep learning and attention processes. In the future, adopting a comprehensive strategy that integrates technical improvements with clinical findings will be crucial. With more study, validation, and cooperation, we aspire that models like ours may be effectively incorporated into healthcare systems, serving as indispensable resources for physicians globally.

\section{acknowledgment}
Mostafa Karami is currently a graduate assistant at the University of Connecticut's Computer Science and Engineering Department.


\vspace{12pt}

\end{document}